# Multigroup Multicast Precoding in Massive MIMO

Meysam Sadeghi[*][†], Emil Björnson[†], Erik G. Larsson[†], Chau Yuen[*], and Thomas L. Marzetta[‡]

[*] Singapore University of Technology and Design, Singapore

[†] Department of Electrical Engineering (ISY), Linköping University, Linköping, Sweden

[‡] Department of Electrical and Computer Engineering, New York University, New York, USA

Email: meysam@mymail.sutd.edu.sg, emil.bjornson@liu.se, erik.g.larsson@liu.se, yuenchau@sutd.edu.sg, and tom.marzetta@nyu.edu

*Abstract*—Optimal physical layer multicasting (PLM) is an NP-hard problem that for simplicity has been studied under idealistic assumptions, e.g., availability of perfect channel state information (CSI), both at the base station (BS) and at the user terminals (UTs). With the advent of massive multiple-input-multiple-output (MIMO), PLM has become more challenging, as the computational complexity of the precoder design is proportional to the number of BS antennas. In this paper, we address these issues by introducing computationally efficient precoders that account for practical CSI acquisition. We derive achievable spectral efficiency expressions for the proposed precoders. Then we introduce a novel problem formulation for the max-min fairness power control that accounts the CSI acquisition overhead, uplink training and downlink transmission powers. We solve this problem and find the optimal uplink and downlink power control policies in closed form. Using numerical simulations, we verify the effectiveness of our proposed schemes compared to the-state-of-the-art PLM schemes for massive MIMO systems.

*Index Terms*—Physical layer multicasting, Massive MIMO, beamforming, large-scale antenna systems.

## I. INTRODUCTION

The mobile data traffic has increased in an unprecedented manner during the last decade. It is anticipated that the mobile data traffic will reach 49 exabytes per month by 2021 [2]. A portion of this huge mobile data traffic belongs to the digital contents that are of interest for groups of users rather than a particular user, e.g., broadcast of sporting events, live shows, news notifications, and etc. To enable an efficient delivery of such traffic, different standards have been introduced, e.g., eMBMS by the 3rd Generation Partnership Project [3] and DVB-H by the digital video broadcasting project [4]. The core concept of these standards is to transmit data from a single transmitter to multiple receivers [3], [4].

The same target, efficient transmission of common messages to groups of users, has also been pursued in academia under the name of physical layer multicasting (PLM) [5], [6]. Particularly two classes of problems have been studied: quality of service (QoS) and max-min fairness (MMF) problems. Utilizing the channel state information (CSI) at the transmitter, the former problem designs precoding vectors that minimize the power consumption at the transmitter while guaranteeing a prescribed quality of service for the user terminals (UTs) [6]. The latter, using CSI at the transmitter, designs precoding

This work was supported in part by the Swedish Research Council (VR), the Swedish Foundation for Strategic Research (SSF), and ELLIIT. Also, it was supported by A*Star SERC project number 142-02-00043. The first author was visiting Linköping University when the work was performed.

A more comprehensive version of this work has been submitted to IEEE Transactions on Wireless Communications [1].

vectors that maximize the minimum value of a quality of service metric, given a fixed power budget [6].

PLM can further be categorized based on the number of multicast messages. If we have multiple distinct multicast messages and multiple groups of UTs such that the UTs in each group are interested (subscribed) to the same message and regard other multicast messages as interference, it is called multigroup PLM [6]. As a special case, if we have a single multicast message, it is called single-group PLM [5].

A first seminal work on PLM is [5], where a single-group single-cell PLM system is studied. It is shown that both QoS and MMF problems are NP-hard. Also, sub-optimal solutions for both problems are presented based on a semidefinite relaxation (SDR) technique. However, the extension of [5] to multigroup PLM is more involved, as the SINR of every UT is affected by the precoding vectors of all multicasting groups. The multigroup single-cell PLM is studied in [6], where it uses SDR technique with randomization and multigroup multicast power control.

The aforementioned works are based on the SDR technique, which has a high computational complexity. Considering a multigroup PLM system with an $N$-antenna BS that serves $K_{tot}$ UTs in $G$ multicasting groups, then solving the QoS problem by SDR method requires $O(\sqrt{GN})$ iterations of an interior point method that each iteration has $O(G^3N^6 + K_{tot}GN^2)$ arithmetic operations [6]. Such a huge computational complexity prevents SDR techniques to be applied in large dimensional systems, e.g. massive MIMO [7].

As massive MIMO systems provide high energy and spectral efficiencies [7]–[12], recent works on multicasting have tried to address the computational complexity of massive MIMO multicasting [13]–[15]. Particularly, [13] presents a successive convex approximation technique for single-group single-cell multicasting of large-scale antenna arrays which reduces the computational complexity to $O(N^{3.5})$. The system set-up of [13] has been extended to a multi-group single-cell multicasting in [14]. Therein a feasible point pursuit based algorithm with a complexity of $O((GN)^{3.5})$ is presented. However, the complexity is still high for large-scale antenna systems. Recently a low computational complexity algorithm, $O(N)$ for single-group and $O(GN^2)$ for multi-group multicasting, for massive MIMO system is presented in [15]. [15] not only significantly reduces the computational complexity but also outperforms the SDR based methods [5], [6].

The common denominator of the aforementioned algorithms is the perfect CSI assumption, both at the BS and at the UTs. However, in practice perfect CSI is not available neither at the BS nor at the UTs, and should be obtained by training. This introduces new challenges to the multicasting problem, which is already NP-hard. To address the CSI acquisition problem, two approaches have been presented in the literature. The first approach leverages the asymptotic orthogonality of the channels in massive MIMO, which simplifies the precoding design [16], [17]. The main problem with the asymptotic approach is that a very large number of antennas, e.g., $N > 1000$, is required to get close to the asymptotic performance, while the performance is poor for realistic antenna numbers [16], [17]. The second approach relies on employing predefined multicasting precoders [18]. More precisely, considering a single-cell multi-group multicasting system, [18] presents a maximum ratio transmission (MRT) based multicast precoder with a novel pilot allocation strategy.

In this paper, we present two different precoding schemes for max-min fair transmit beamforming in massive MIMO multicasting systems. The first scheme is a two-layered precoder, inspired from [15], that employs multicast transmission.[1] The second scheme uses a unicast transmission[1] and is a ZF precoder optimized for massive MIMO multicasting. Particularly, we present the following contributions:

- We present computationally efficient precoders for multicasting in massive MIMO systems while accounting for imperfect CSI, at the BS and also at the UTs.
- Considering the proposed precoders, we derive closed-form expressions for the achievable SE of each UT.
- We introduce a novel framework for the MMF problem and then solve it for the proposed precoders. We present the optimal uplink training and downlink transmission power policies, all in closed-form.
- We compare our results with the state of the art algorithms and validate their effectiveness numerically.

The rest of the paper is organized as follows. Section II introduces the system and signal model. Section III presents the proposed precoders and the derived achievable SEs. Section IV studies the MMF problem and presents the optimal uplink training and downlink power policies. Section V presents the numerical results. Finally, Section VI concludes the paper.

## II. SYSTEM AND SIGNAL MODEL

We consider multi-group multicasting in a single-cell massive MIMO system.[2] We assume the system has one BS with $N$ antennas and it transmits $G$ data streams toward $G$ multicasting groups. We denote the set of indices of these $G$ multicasting groups as $\mathcal{G}$, i.e., $\mathcal{G} = \{1, \ldots, G\}$. We assume the $j$th data stream, $j \in \{1, \ldots, G\}$, is of interest for $K_j$ UTs, and we say that these $K_j$ UTs belong to the $j$th multicasting group. We denote the set of indices of all the UTs in the $j$th multicasting

---

group as $\mathcal{K}_j$, i.e. $\mathcal{K}_j = \{1, \ldots, K_j\}$. Therefore $|\mathcal{G}| = G$ and $|\mathcal{K}_j| = K_j$. We assume each UT is assigned to just one multicasting group. We denote the total number of UTs in the system as $K_{tot} = \sum_{j=1}^{G} K_j$.

We consider a block flat-fading channel model where $C_B$ (in Hz) is the coherence bandwidth and $C_T$ (in seconds) is the coherence time. Hence the channels are static within a coherence interval of $T = C_B C_T$ symbols. We assume the BS does not have a priori CSI of the UTs but estimates the channel by uplink pilots transmission using a TDD protocol, exploiting channel-reciprocity. The procedure is detailed below. Under these conventions, we represent the channel between the BS and UT $k$ in multicasting group $j$ as $\mathbf{g}_{jk}$. We assume all the UTs have independent Rayleigh fading channels [19]. This implies that $\mathbf{g}_{jk} \sim \mathcal{CN}(\mathbf{0}, \beta_{jk}\mathbf{I}_N)$, $\forall k, j$, where $\beta_{jk}$ represents the large scale fading, which are strictly positive and known at the BS [7].

### A. Channel Estimation

The BS uses uplink pilot transmission to estimate the channels to the UTs in the system. It can be performed by using one orthogonal pilot per UT (so it requires $K_{tot}$ pilots). Denoting the pilot length as $\tau_p$, to have orthogonal pilots we need $\tau_p \geq K_{tot}$. The MMSE estimate of the channel of UT $k$ in group $j$ is

$$\hat{\mathbf{g}}_{jk} = \frac{\sqrt{\tau_p p_{jk}^u} \beta_{jk}}{1 + \tau_p p_{jk}^u \beta_{jk}} \left( \sqrt{\tau_p p_{jk}^u} \mathbf{g}_{jk} + \mathbf{n} \right) \quad (1)$$

where $\mathbf{n} \sim \mathcal{CN}(\mathbf{0}, \mathbf{I}_N)$ is normalized additive noise and $p_{jk}^u$ is the uplink pilot power of UT $k$ in group $j$. Therefore we have $\hat{\mathbf{g}}_{jk} \sim \mathcal{CN}(\mathbf{0}, \gamma_{jk}\mathbf{I}_N)$ with $\gamma_{jk} = (\tau_p p_{jk}^u \beta_{jk}^2)/(1 + \tau_p p_{jk}^u \beta_{jk})$. The independent estimation error is $\tilde{\mathbf{g}}_{jk} = \hat{\mathbf{g}}_{jk} - \mathbf{g}_{jk} \sim \mathcal{CN}(\mathbf{0}, (\beta_{jk} - \gamma_{jk})\mathbf{I}_N)$. Moreover, we denote the $N \times K_{tot}$ matrix obtained by stacking the estimated channel of all the UTs in the system as $\hat{\mathbf{G}} = [\hat{\mathbf{G}}_1, \ldots, \hat{\mathbf{G}}_G]$, where $\hat{\mathbf{G}}_j = [\hat{\mathbf{g}}_{j1}, \ldots, \hat{\mathbf{g}}_{jK_j}] \; \forall j \in \mathcal{G}$.

### B. Unicast versus Multicast Transmission

Let us denote by $s_i \sim \mathcal{CN}(0, 1) \; \forall i \in \mathcal{G}$ the signal requested by the UTs in the $i$th multicasting group, i.e., $\mathcal{K}_i$. Note that $s_i$ is independent across $i$. We stack them in a vector $\mathbf{s} = [s_1, \ldots, s_G]^T$. In multicast transmission, we use $\mathbf{s}$ as the data vector. Also the precoding matrix becomes an $N \times G$ matrix $\mathbf{W} = [\mathbf{w}_1, \ldots, \mathbf{w}_G]$ where $\mathbf{w}_j$ is the joint precoding vector of all the UTs in $j$th multicasting group. The received signal of UT $k$ in multicasting group $j$ is

$$y_{jk} = \mathbf{g}_{jk}^H \mathbf{W} \mathbf{s} + n = \mathbf{g}_{jk}^H \sum_{i=1}^{G} \mathbf{w}_i s_i + n. \quad (2)$$

For unicast transmission the transmit data vector is a $K_{tot} \times 1$ vector

$$\mathbf{x} = [\underbrace{s_1, \ldots, s_1}_{\text{group 1}}, \underbrace{\cdots}_{\text{groups 2 to G-1}}, \underbrace{s_G, \ldots, s_G}_{\text{group G}}]^T \quad (3)$$

and the precoding matrix is an $N \times K_{tot}$ matrix $\mathbf{V}$. Therefore, the received signal of UT $k$ in multicasting group $j$ is

$$y_{jk} = \mathbf{g}_{jk}^H \mathbf{V}\mathbf{x} + n = \mathbf{g}_{jk}^H \sum_{i=1}^{G} \sum_{t=1}^{K_i} \mathbf{v}_{it} s_i + n \quad (4)$$

where $\mathbf{v}_{it}$ is the precoding vector of UT $t$ in group $i$. Note that for unicast transmission we have a dedicated precoding vector for each UT in every multicast group.

## III. THE PRECODERS AND ACHIEVABLE SEs

In this section, we present two precoding schemes for multicasting in massive MIMO systems. The first scheme is a two-layered precoder, $\mathbf{W}$, that employs multicast transmission. The second scheme is simply a ZF precoder, $\mathbf{V}$, that uses unicast transmission to perform multicasting. More precisely, it considers the UTs within each multicast group as if they are requesting different messages and performs ordinary unicast transmission. For each of these schemes, we derive the achievable SE of each UT in the system, while accounting for the lack of a priori CSI, at both BS and UT sides.

### A. The Two-layered Multicast Precoder

In [15], a computationally efficient precoder is proposed for multigroup massive MIMO multicasting, that significantly outperforms SDR-based methods. It employs a two-layered precoder, where the outer layer cancels the inter group interference and the inner layer is designed to maximize the minimum SINR of the system while satisfying the power constraint at the BS. However, it assumes 1) perfect CSI is available at both BS and UTs, and 2) focuses on instantaneous SINR rather than SE. These simplifying assumptions have also been used in the literature, e.g., [5], [6], [13]–[15], however 1) channels are unknown and need to be estimated, and 2) as SE accounts for the CSI acquisition overhead, it is more practical to consider the SE than the SINR as the performance metric.

Here, we present a two-layered precoding scheme based on [15] and then derive an achievable SE that addresses both of the aforementioned issues. Our proposed precoding matrix is an $N \times G$ matrix $\mathbf{W} = [\mathbf{w}_1, \ldots, \mathbf{w}_G]$ where $\mathbf{w}_j$ is the precoding vector for $j$th multicast group and

$$\mathbf{w}_j = \underbrace{\left( \mathbf{I}_N - \hat{\mathbf{G}}_{-j} \left( \hat{\mathbf{G}}_{-j}^H \hat{\mathbf{G}}_{-j} \right)^{-1} \hat{\mathbf{G}}_{-j}^H \right)}_{\text{outer layer}} \underbrace{\sum_{k=1}^{K_j} \sqrt{\mu_{jk}} \hat{\mathbf{g}}_{jk}}_{\text{inner layer}} \quad (5)$$

where $p_{jk}^{dl}$ is the downlink power of UT $k$ in group $j$, $\hat{\mathbf{G}}_{-j} = [\hat{\mathbf{G}}_1, \ldots, \hat{\mathbf{G}}_{j-1}, \hat{\mathbf{G}}_{j+1}, \ldots, \hat{\mathbf{G}}_G]$, and $\mu_{jk} = \sqrt{p_{jk}^{dl} / ((N - \nu_j)\gamma_{jk})}$ with $\nu_j = K_{tot} - K_j$.

The outer layer of $\mathbf{w}_j$ cancels the inter group interference, within the limitations of the channel estimation errors. The inner layer is designed to optimize the MMF problem, which is detailed next. Notice that $\mathbb{E}[\|\mathbf{w}_j\|^2] = \sum_{k=1}^{K_j} p_{jk}^{dl}$ and the total power used at the BS becomes $P = \sum_{j=1}^{G} \sum_{k=1}^{K_j} p_{jk}^{dl}$.

Considering (5), as $\hat{\mathbf{G}}_{-j}$ is an $N \times \nu_j$ matrix, the computational complexity of calculating $\mathbf{w}_j$ is $O(N^2 \nu_j + N \nu_j^2 + \nu_j^3)$, which is significantly less than its counterparts, e.g., the SDR-based algorithms for multicasting have a complexity of $O(N^{6.5} G^{3.5} + N^{2.5} G^{1.5} K_{tot})$ [6]. Therefore, (5) presents a relatively computationally efficient method for calculating the precoders.

Now we need to obtain a measure of the performance of (5). We have the following result on the achievable SE of the proposed precoder.

**Theorem 1.** *Employing the proposed two-layered multicast precoder, i.e.,* (5)*, an achievable SE for user* $k$ *of group* $i$ *is*

$$\text{SE}_{ik}^{mu} = \left( 1 - \frac{\tau_p}{T} \right) \log_2(1 + \text{SINR}_{ik}^{mu}). \quad (6)$$

*where*

$$\text{SINR}_{ik}^{mu} = \frac{(N - \nu_i)\gamma_{ik} p_{ik}^{dl}}{1 + \gamma_{ik} \sum_{m=1}^{K_i} p_{im}^{dl} + (\beta_{ik} - \gamma_{ik})P} \quad (7)$$

*is the effective SINR of this user.*

*Proof Sketch.* Start from (2) and replace $\mathbf{g}_{jk}$ with $\hat{\mathbf{g}}_{jk} - \tilde{\mathbf{g}}_{jk}$. Apply the use-and-then-forget bounding technique [12], while noting that the outer layer of (5) cancels the inter group interference, within the limitations of the channel estimation errors, and the remaining interference is uncorrelated with the desired signal term. □

### B. The ZF Unicast Precoder

Consider an $N \times K_{tot}$ precoding matrix $\mathbf{V} = [\mathbf{v}_{11}, \ldots, \mathbf{v}_{jk}, \ldots, \mathbf{v}_{GK_G}]$ for unicast transmission. Then the ZF precoding vector of UT $k$ in group $j$ is

$$\mathbf{v}_{jk} = \sqrt{p_{jk}^{dl} \gamma_{jk}(N - K_{tot})} \; \hat{\mathbf{G}}(\hat{\mathbf{G}}^H \hat{\mathbf{G}})^{-1} \mathbf{e}_{\nu_{jk}, K_{tot}} \quad (8)$$

where $\nu_{jk} = \sum_{t=1}^{j-1} K_t + k$ and $p_{jk}^{dl}$ is the downlink power allocated to this user. Note that $\mathbb{E}[\|\mathbf{v}_{jk}\|^2] = p_{jk}^{dl}$ and we denote the total utilized downlink power as $P = \sum_{j=1}^{G} \sum_{k=1}^{K_j} p_{jk}^{dl}$.

Considering (8), as $\hat{\mathbf{G}}$ is an $N \times K_{tot}$ matrix, the computational complexity of calculating $\mathbf{v}_{jk}$ is $O(NK_{tot}^2 + K_{tot}^3)$. Given (8), we can achieve the following SEs for the UTs in the system.

**Proposition 1.** *Employing the proposed ZF unicast precoder, i.e.,* (8) *, an achievable SE for user* $k$ *of group* $i$ *is*

$$\text{SE}_{ik}^{un} = \left( 1 - \frac{\tau_p}{T} \right) \log_2(1 + \text{SINR}_{ik}^{un}) \quad (9)$$

*where*

$$\text{SINR}_{ik}^{un} = \frac{(N - K_{tot})\gamma_{ik} p_{ik}^{dl}}{1 + (\beta_{ik} - \gamma_{ik})P} \quad (10)$$

*is the effective SINR of this user.*

*Proof.* The proof follows conventional bounding technique as [12] and is omitted for brevity. □

We will now take a closer look at the structure of the achievable SINRs under the two proposed schemes, i.e., (7) and (10). We have an array gain of $N - \nu_i$ or $N - K_{tot}$, respectively in the numerators of (7) and (10). Also, the interference terms in their denominators have a component of the form $(\beta_{ik} - \gamma_{ik})P$, due to ZF-based transmission. Also note that the higher array gain in the numerator of (7) compared to (10), is due to the elimination of the intra-group interference cancellation in the outer layer of (5), which in turn has resulted in an extra interference term in the denominator of (7), i.e., $\gamma_{ik} \sum_{m=1}^{K_i} p_{im}^{dl}$.

## IV. Max-Min Fairness Power Control

In this section, we introduce the MMF power control problem and derive the optimal uplink training power and downlink transmit power policies in closed form. Assuming perfect CSI, the existing works in the literature [5], [6], [14]–[18], [20] consider the instantaneous SINR as the metric of interest and just account for the available power at the BS, while ignoring CSI acquisition and its associated overhead. Here we introduce a more general problem formulation for the MMF problem that accounts for 1) the CSI acquisition, 2) choose the SE as the metric of interest, and 3) expand the set of resource constraints covering 3-a) the available power at the BS, 3-b) the uplink training power limit of the UTs, and 3-c) the length of the uplink training pilots. The proposed MMF problem is

$$\mathcal{P}1: \max_{\tau_p, \{p_{jk}^{dl}\}, \{p_{jk}^u\}} \min_{\forall j \in \mathcal{G}} \min_{\forall k \in \mathcal{K}_j} (1 - \frac{\tau_p}{T}) \log_2(1 + \text{SINR}_{jk}^{\dagger}) \quad (11)$$

$$s.t. \quad p_{jk}^u \leq p_{jk}^{utot} \quad \forall\, k \in \mathcal{K}_j, \forall\, j \in \mathcal{G} \quad (11\text{-}C1)$$

$$P = \sum_{j=1}^{G} \sum_{k=1}^{K_j} p_{jk}^{dl} \leq P_{tot} \quad (11\text{-}C2)$$

$$\tau_p \in \{K_{tot}, \ldots, T\} \quad (11\text{-}C3)$$

where $\text{SINR}_{jk}^{\dagger}$ is either $\text{SINR}_{jk}^{mu}$ or $\text{SINR}_{jk}^{un}$, $p_{jk}^{utot}$ is the maximum pilot power limit of user $k$ in group $j$, and $P_{tot}$ is the total available power at the BS. The constraint (11-C2) is due to the total available power at the BS.

It is straightforward to show that in $\mathcal{P}1$, constraint (11-C2) should be met with equality. To see this, assume the contrary, e.g., at the optimal solution of $\mathcal{P}1$ we have $P_{tot} > P = \sum_{j=1}^{G} \sum_{k=1}^{K_j} p_{jk}^{dl}$. Then one can increase all the $p_{jk}^{dl}$ by a factor of $\frac{P_{tot}}{P}$. This increases the SE of each UT, hence improves the minimum SE of the system, which contradicts our assumption. Hence at the optimal solution of $\mathcal{P}1$, $P = P_{tot}$.

To solve $\mathcal{P}1$, we use a two-step approach. First, we solve it for a fixed value of $\tau_p$ and determine its optimal solution in closed form. Second, we find the optimal value of $\tau_p$ by searching over all $T - K_{tot} + 1$ different values, thanks to the closed-form expression obtained in the first step. Given an arbitrary $\tau_p$, as the logarithm is a strictly increasing function, $\mathcal{P}1$ can be replaced with a problem $\mathcal{P}'1$ as follows

$$\mathcal{P}'1: \max_{\{p_{jk}^u\}, \{p_{jk}^{dl}\}} \min_{\forall j \in \mathcal{G}} \min_{\forall k \in \mathcal{K}_j} \text{SINR}_{jk}^{\dagger} \quad (12)$$

$$s.t. \quad (11\text{-}C1) \quad (12\text{-}C1)$$

$$P_{tot} = P. \quad (12\text{-}C2)$$

Now we have the following results for $\mathcal{P}'1$.

**Theorem 2.** *Considering the proposed precoder in* (5)*, at the optimal solution of $\mathcal{P}'1$ all the UTs within the system will*

*receive the same effective SINR, i.e., $\Gamma = \text{SINR}_{ik} \; \forall i, k$, and it is the solution of the equation*

$$P = \sum_{i=1}^{G} \frac{\Gamma \Delta_i}{N - \nu_i - \Gamma K_i} \quad (13)$$

*where*

$$\Delta_i = \sum_{k=1}^{K_i} (\frac{1}{\gamma_{ik}^*} + P \frac{\beta_{ik}}{\gamma_{ik}^*} - P) \quad (14)$$

$$\gamma_{ik}^* = \frac{\tau_p p_{ik}^{utot} \beta_{ik}^2}{1 + \tau_p p_{ik}^{utot} \beta_{ik}}. \quad (15)$$

*The optimal uplink training and downlink transmission powers of UT $k$ in group $i$ are*

$$p_{ik}^{u*} = p_{ik}^{utot} \quad (16)$$

$$p_{ik}^{dl*} = \frac{\Gamma}{N - \nu_i} \left( \frac{1}{\gamma_{ik}^*} + P_i^{dl} + P \frac{\beta_{ik}}{\gamma_{ik}^*} - P \right) \quad (17)$$

*where $P_i^{dl} = \frac{\Gamma \Delta_i}{N - \nu_i - \Gamma K_i}$.*

*Proof Sketch.* Starting from $\text{SINR}_{ik}^{mu}$, given in Theorem 1, $\text{SINR}_{ik}^{mu}$ and $\gamma_{ik}$ monotonically increase with respect to $p_{ik}^u$. Therefore, $\gamma_{ik}$ and $p_{ik}^u$ respectively follow (15) and (16). Then it should be proved that at the optimal solution $\text{SINR}_{ik} = \text{SINR}_{jt} = \Gamma \; \forall k, t, i, j$, where $\Gamma$ is fixed. Then we find $p_{ik}^{dl*}$ based on $\Gamma$. Now summing up over all $p_{ik}^{dl*}$ and employing (12-C2) one can obtain the desired result. □

**Theorem 3.** *Considering the proposed precoder in* (8)*, at the optimal solution of $\mathcal{P}'1$ all the UTs within the system will receive the same SINR, i.e., $\Gamma = \text{SINR}_{ik} \; \forall i, k$, and it is equal to*

$$\Gamma = \frac{(N - K_{tot})P}{\sum_{i=1}^{G} \sum_{k=1}^{K_i} \frac{1 + (\beta_{ik} - \gamma_{ik}^*)P}{\gamma_{ik}^*}} \quad (18)$$

*where $\gamma_{ik}^*$ is given in* (15)*. The optimal uplink training and downlink transmission powers of UT $k$ in group $i$ are*

$$p_{ik}^{u*} = p_{ik}^{utot} \quad (19)$$

$$p_{ik}^{dl*} = \frac{1 + (\beta_{ik} - \gamma_{ik}^*)P}{\gamma_{ik}^*(N - K_{tot})} \Gamma. \quad (20)$$

*Proof Sketch.* First it should be shown that for every UT $k$ in group $i$ its SINR increases monotonically with $p_{ik}^u$, which results in (19). Then it should be shown that at the optimal solution all UTs will have the same SINR, which also determines (20). Now using this fixed value for the SINR and the downlink transmission power constraint, (12-C2), one can obtain the desired result. □

Theorems 2 and 3 determine the optimal value of the effective SINR, uplink training powers, and downlink transmission powers in closed form, for any given pilot length. These closed-form expressions enable us to find the optimal value of SE by simply searching over $\tau_p \in \{K_{tot}, \ldots, T\}$ and find the pilot length that provides the highest SE. This method has been used in our numerical simulations.

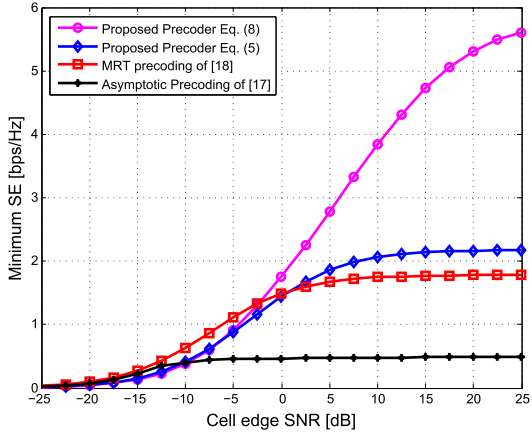

Fig. 1: SE versus cell edge SNR.

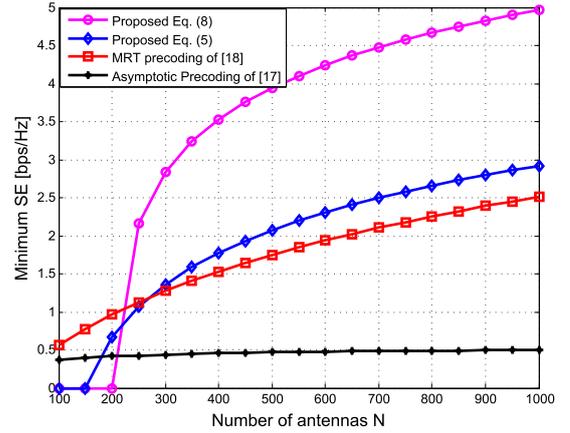

Fig. 2: SE versus number of BS antennas N, high SNR regime.

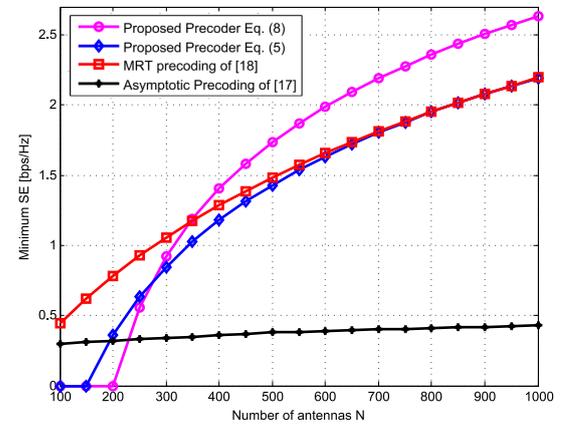

Fig. 3: SE versus number of BS antennas N, low SNR regime.

## V. NUMERICAL RESULTS

In this section, we use numerical simulations to assess the performance of our proposed precoders. We compare them with the-state-of-the-art massive MIMO multicast precoders proposed in [17] and [18]. [18] presents an MRT precoder that accounts for CSI acquisition at the BS and derives an achievable SE without CSI at the UTs, exploiting the channel hardening, and [17] uses an asymptotic approach to design the precoders. In our simulations, we consider a system with $G$ multicasting groups where each group has $K$ UTs, i.e., $K_i = K \; \forall i \in \mathcal{G}$. The cell radius is considered to be 500 meters and the UTs are randomly and uniformly distributed in the cell excluding an inner circle of radius 35 meters. The large-scale fading parameters are modeled as $\beta_{ik} = \bar{d}/x_{ik}^\nu$ where $\nu = 3.76$ is the path-loss exponent and the constant $\bar{d} = 10^{-3.53}$ regulates the channel attenuation at 35 meters [21]. Also $x_{ik}$ is the distance between UT $k$ in group $i$ and the BS in meters. At a carrier frequency of 2 GHz, the transmission bandwidth (BW) is assumed to be 20 MHz, the coherence bandwidth and coherence time are considered to be 300 kHz and 2.5 ms, which results in a coherence interval of length 750 symbols for a vehicular system with speed of 108 kilometers per hour [12]. The noise spectral density is considered to be −174 dBm/Hz.

As shown in Theorems 1 and 2, the SE depends on the system parameters $N$, $G$, $K$, and the SNR of the UTs. Fig. 1 presents the effect of the SNR on the minimum SE of the system. Therein, we consider a massive MIMO multicasting system with $N = 500$ antennas at the BS, $G = 4$ multicast groups, and $K = 50$ UTs per group, where the cell edge SNR is changing from −25 dB to 25 dB. As seen, when the cell edge SNR is greater than −2.5 dB, our proposed ZF unicast precoder provides higher SE than the other schemes, and for SNRs less than −2.5 dB [18] has higher SE. This is due to the fact that our proposed precoders have structures similar to ZF while [18] is based on MRT precoding.

Fig. 2 presents the minimum SE for different numbers of BS antennas, $N$. The cell edge SNR is set to 10 dB, $G = 4$, and $K = 50$. Note that by adding more antennas the SE of all schemes improves. However, the SEs of our proposed precoders improve faster, compared to [17] and [18], and when we have more than 275 antennas, both of them provide higher SEs than [17] and [18]. This makes it appealing for large-scale antenna systems. Also note that our proposed precoders and [18], provide significantly higher SE than the asymptotic precoding approach.

As the SNR affects the performance of the proposed precoders, Fig. 3 presents the minimum SE of the system for different number of BS antennas for low SNR regime, i.e., cell edge SNR of 0 dB. Note that the same observations as in Fig. 2 holds for Fig. 3. But now our ZF unicast precoder (two layered multicast precoder) requires $N > 350$ ($N > 600$) antennas, to outperform [18].

In PLM we might have a large number of multicasting groups or UTs in the system. Fig. 4 presents the minimum SE of the system versus $G$, where we have $K = 20$ UTs in each multicast group and the BS has $N = 300$ antennas. When $G$ (or $K_{tot}$) is small, the proposed ZF unicast precoder has the highest SE and as we increase $G$ (or $K_{tot}$), the MRT precoder of [18] becomes the superior scheme. This is due to the fact that the proposed ZF unicast scheme uses more time-frequency resources to achieve accurate channel estimates. Conversely, the MRT precoder of [18] uses a common pilot for all the UTs in each multicast group and creates deliberate

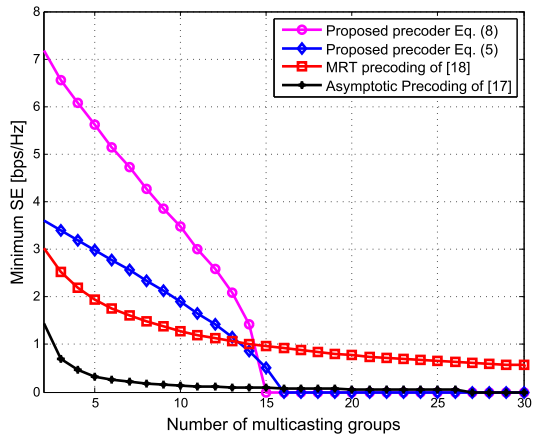

Fig. 4: SE versus number of multicast groups G.

pilot contamination to save time-frequency resources.

Based on the above discussions one can observe that the optimal precoding scheme depends on the system parameters, e.g., $N$, $G$, $K$, and is either our proposed unicast ZF precoder or the one presented in [18]. Therefore a desirable massive MIMO multicasting system should be able to switch between these two schemes based on the system parameters. Note that the switching policy can easily be implemented as the achievable SEs under these two schemes have closed-form expressions, which are functions of system parameters that are changing slowly, e.g., large-scale fading coefficients. This is in sharp contrast with existing schemes which are based on system parameters that are changing frequently, e.g., small-scale fading. Moreover, in all of the simulations, our proposed schemes account for CSI acquisition at the BS by channel estimation and did not require instantaneous CSI at the UTs due to the channel hardening, which makes them desirable for practical massive MIMO multicasting systems. In contrast, the existing approaches, for example [5], [6], [13]–[15], are based on the assumptions of availability of perfect CSI at the BS and UTs, which is impractical.

## VI. CONCLUSION

We studied physical layer multicasting in massive MIMO systems. Contrary to most existing works, which are based on idealistic assumptions or analysis, e.g., perfect CSI or asymptotic analysis, our proposed precoders introduced practical methods for physical layer multicasting in massive MIMO systems. We presented not only computationally efficient precoders, but also we addressed the CSI availability at both BS and UTs leveraging uplink channel estimation and channel hardening, respectively. We showed that our proposed precoders significantly outperform the asymptotic precoding approach [17]. Moreover, we revealed that a desired multigroup physical layer multicasting massive MIMO system should support two precoding schemes, our proposed ZF unicast precoder and the MRT scheme from [18], and switch between them according to the system parameters. Moreover, the switching policy can be easily implemented, thanks to the derived closed-form expressions for the SE of the system. Finally, we note that a much more comprehensive study of multigroup multicasting is available in [1], which is an extended version of this paper.